\documentclass[preprint,floatfix] {revtex4} 
\newcommand{\rvec}{\mathrm {\mathbf {r}}} 
\newcommand{\pvec}{\mathrm {\mathbf {p}}} 
\usepackage{graphicx}
\usepackage{subfigure}
\usepackage{xcolor}
\usepackage{amsmath}
\usepackage{enumerate}
\usepackage{longtable}
\usepackage{tabularx}

\usepackage{color, soul}
\definecolor{darkblue}{rgb}{0,0,0.5}
\setulcolor{darkblue}

\begin{document}

\title{Fisher information in confined isotropic harmonic oscillator}

\author{Neetik Mukherjee}
%%\altaffiliation{Email: neetik.mukherjee@iiserkol.ac.in.}

\author{Amlan K.~Roy}
\altaffiliation{Corresponding author. Email: akroy@iiserkol.ac.in, akroy6k@gmail.com.}
\affiliation{Department of Chemical Sciences\\
Indian Institute of Science Education and Research (IISER) Kolkata, 
Mohanpur-741246, Nadia, WB, India}

\begin{abstract}
%%1234567890 %%1234567890 %%1234567890 %%1234567890 %%1234567890 %%1234567890 %%1234567890 %%1234567890 %%1234567890 %%1234567890

Fisher information ($I$) is investigated in a confined harmonic oscillator (CHO) enclosed in a spherical enclosure, in conjugate 
$r$ and $p$ spaces. A comparative study between CHO and a free quantum particle in spherical box (PISB), as well as CHO and 
respective free harmonic oscillator (FHO) is pursued with respect to energy spectrum and $I$. This reveals that, a CHO offers 
two exactly solvable limits, namely, a PISB and an FHO. Moreover, the dependence of $I$ on quantum numbers $n_{r}, l, m$ in FHO 
and CHO are 
analogous. The role of force constant is discussed. Further, a thorough systematic analysis of $I$ with respect to variation of 
confinement radius $r_c$ is presented, with particular attention on \emph{non-zero}-$(l,m)$ states. Considerable new important
observations are recorded. The results are quite accurate and most of these are presented for the first time. 

\vspace{5mm}

\vspace{5mm}

\end{abstract}
\maketitle

\section{introduction}

Confined quantum systems have generated appreciable interest in the area of physics, chemistry, biology, \cite{michels37,
sabin2009,sen12,sen2014electronic}, etc., since the first decade of this century. The simplest model system, namely, a particle 
in a $D$-dimensional spherical box, is used in textbooks to introduce fundamental ideas of quantum physics like the effect of 
boundary conditions, quantization of energy, linear superposition of wave etc. In such situations, the particle shows 
fascinating distinguishable changes in its observable physical, chemical properties \cite{aquino16,yu17}. These models 
were designed to mimic real physical and chemical systems. In past two decades, behavior of confined systems with 
different potentials were explored quite vigorously. It has witnessed profound applications in a variety of research areas,
like condensed matter, semiconductor physics, astrophysics, nano-science and technology, quantum dots, wires and wells 
\cite{sabin2009,sen2014electronic,pang11}, etc. Atomic, molecular confinement either in fullerene cage or inside the cavities 
of zeolite molecular sieves \cite{sabin2009,sen12,sen2014electronic}, were also pursued with great enthusiasm and promise.

A 1D confined harmonic oscillator (CHO) is considered as an intermediate model between particle in a box and free harmonic 
oscillator (FHO) \cite{Gueorguiev06,laguna14}. A variety of theoretical methods \cite{navarro80,taseli93,campoy02,montgomery10,
roy15}, such as semiclassical WKB, power-series solution, Pad\'e approximation, perturbation theory, hypervirial theorem, 
Rayleigh-Ritz variation method, finding the zeros of confluent hypergeometric function numerically, imaginary-time evolution 
method etc., have been employed. Apart from the eigenspectrum, considerable efforts were directed towards Einstein co-efficients
as well as information measures \cite{barton90, berman91,tanner91,aquino01,steva10,laguna14,ghosal16} etc. Its 3D counterpart, 
i.e., the isotropic CHO centrally enclosed in a spherical box with impenetrable walls has been treated with equal intensity, 
using a number of theoretical methods, such as WKB, supersymmetric, properties of hypergeometric functions, quantization rule, 
generalized pseudospectral (GPS) method \cite{fernandez81,navarro83,aquino97,sen06,montgomery07,steva08,montgomery10,roy14}. It 
exhibits several unique degeneracy, splitting pattern as well as symmetry breaking, related to \emph{simultaneous, incidental 
and inter-dimensional} degeneracy. 

Over the years information theory has flourished as a pertinent field of study to revisit quantum mechanical problems. 
Apparently, this theory deals with single-particle probability density $\rho(\tau)$ of a system. In reality, information 
measures like R\`enyi ($R$) and Shannon ($S$) entropy, Fisher information ($I$), Onicescu energy ($E$), quantify the 
spatial delocalization of $\rho(\tau)$, and hence can be explicitly employed in numerous interesting occurrences 
in quantum mechanics \cite{sabin2009,sen12,sen2014electronic}. In a recent work present authors have thoroughly investigated  
all these information theoretic measures ($R,~S,~E,~I$ and complexity) for confined hydrogen atom (CHA) \cite{mukherjee17a,mukherjee17,majumdar17}.  
Fundamentally, $I$ quantifies expected error in a given 
measurement. By definition, it is a gradient functional of density. Because of its sensitivity toward local rearrangement of 
density, it is said to have property of locality \cite{frieden04}. Usually an increase in $I$ suggests localization of 
$\rho(\tau)$. It is akin to the famous Weizs\"acker kinetic energy functional ($T_{\Omega}[\rho]$) frequently employed in 
density functional theory \cite{debajit12}. It is also used in many other context such as to ascertain Pauli 
effects \cite{toranzo14,nagy06a}, ionization potential, polarizability \cite{sen07}, entanglement \cite{nagy06b}, 
avoided crossing \cite{ferez05} in atomic physics, etc., to name a few. Exact analytical form of $I$ in composite $r,~p$-spaces 
for central potential have already been established in \cite{romera05}. In \cite{mukherjee17a,mukherjee17}, authors have demonstrated the same relations
for CHA. Further, $I$ was explored for vibrational 
levels of various diatomic model potentials, such as P\"oschl-Teller \cite{dehesa06}, pseudo-harmonic \cite{yahya14}, 
Tietz-Wei \cite{falaye14}, Frost-Musulin \cite{idiodi16}, Generalized Morse \cite{onate16}, exponential-cosine screened coulomb 
potential \cite{abdelmonem17}, etc.  

The exact analytical form of $I$ in conjugate $r$, $p$ space for 3D FHO was reported in 
\cite{romera05}. Some statistical complexities were investigated for Bohr-like orbits \cite{sanudo08}. In 2016, $S$ and $R$ 
were computed for highly excited quantum states within Laguerre asymptotic approximation \cite{aptekarev16} and some 
linearized method \cite{dehesa16}. However, an elaborate information theoretic study for 3D CHO is still lacking as of now.  
Hence, in this communication our primary objective is to perform a systematic analysis of $I$ in a CHO, for any arbitrary 
state characterized by quantum numbers $n_{r}, l, m$, in both $r$ and $p$ spaces. Special attention has been paid to explore 
the effect of $m$ on $I$, as well as \emph{non-zero l} states. Illustrative calculations are executed with \emph{exact} 
analytical wave functions in $r$-space whereas $p$-space wave functions are obtained from numerical Fourier transform of 
$r$-space counterpart. Specimen results are provided for $1s$-$1g$ as well as $2s$-$2m$ states. All the allowed $m$'s corresponding 
to a given $n_r$ and $l$ have been taken into account, which provides the liberty to follow the definitive transformation in 
properties of states with $m$. Changes are also perceived with respect to oscillator frequency ($\omega$). Since such works 
are very limited, most of present results are reported here for the first time. Section~II gives a brief description about our 
theoretical method used; Sec.~III offers a detailed discussion of results on $I$, while we conclude with a few comments in 
Sec.~IV.
      
\section {Methodology}
The non-relativistic radial Schr\"odinger equation for an isotropic CHO, without any loss of generality, 
may be written as (atomic unit employed, unless otherwise mentioned), 
\begin{equation}
\left[-\frac{1}{2} \ \frac{d^2}{dr^2} + \frac{\ell (\ell+1)} {2r^2} + v(r) +v_c (r) \right] \psi_{n_{r},\ell}(r)=E_{n_{r},\ell}\ 
\psi_{n_{r},\ell}(r),
\end{equation}
where $v(r)=\frac{1}{2}\omega^{2}r^{2}$ and $\omega$ is the oscillator frequency. Our required confinement effect is induced by 
invoking the following form of potential: $v_c(r) = +\infty$ for $r > r_c$, and $0$ for $r \leq r_c$, where $r_c$ denotes 
radius of confinement. 

\emph{Exact} generalized radial wave function for a CHO is mathematically expressed as \cite{montgomery07}, 
\begin{equation}
\psi_{n_{r}, l}(r)= N_{n_{r}, l} \ r^{l} \ _{1}F_{1}\left[\frac{1}{2}\left(l+\frac{3}{2}-
\frac{\mathcal{E}_{n_{r},l}}{\omega}\right),
\left(l+\frac{3}{2}\right),\omega r^{2}\right] e^{-\frac{\omega}{2}r^{2}}.
\end{equation}
Here $N_{n_{r}, l}$ signifies normalization constant, $\mathcal{E}_{n_{r},l}$ corresponds to energy eigenvalue of a given state 
represented by $n_{r},l$ quantum numbers and $_1F_1\left[a,b,r\right]$ is a Kummer confluent hypergeometric function. Energies 
are calculated by assigning Dirichlet boundary condition, $\psi_{n_{r},l} (0)=\psi_{n_{r},l}(r_c)=0$ in Eq.~(2). In this work, 
GPS method has been utilized to generate $\mathcal{E}_{n_{r},l}$ of CHO. Over the years, this method has yielded very accurate 
results for a large number of model and real systems including atoms and molecules; this is well documented in the references 
\cite{roy04,sen06,roy08,roy13,roy14,roy15a} and therein. 

The $p$-space wave function is obtained from Fourier transform of the $r$ counterpart,
\begin{equation}
\begin{aligned}
\psi_{n_{r},l}(p) & = & \frac{1}{(2\pi)^{\frac{3}{2}}} \  \int_0^{r_c} \int_0^\pi \int_0^{2\pi} \psi_{n_{r},l}(r) \ 
\Theta_{l,m}(\theta) \Phi_{m}(\phi) \ e^{ipr \cos \theta}  r^2 \sin \theta \ \mathrm{d}r \mathrm{d} \theta \mathrm{d} \phi .
\end{aligned}
\end{equation}
Here $\psi(p)$ needs to be normalized. The normalized $r$- and $p$-space densities are represented as, 
$\rho(\rvec) = |\psi_{n_{r},l,m}(\rvec)|^2$ and $\Pi(\pvec) = |\psi_{n_{r},l,m} (\pvec)|^2$ respectively. 
Let $I_{\rvec}$, $I_{\pvec}$ denote \emph{net} information measures in consequential $r$ and $p$ space of CHO. It is well
established that, for a single particle in a central potential, these quantities can be written in terms of radial 
expectation values $\langle r^k \rangle $ and $ \langle p^k \rangle (k = -2,2)$ \cite{romera05}, as below, 
\begin{eqnarray} 
I_{\rvec}  =  \int_{{\mathcal{R}}^3} \left[\frac{|\nabla\rho(\rvec)|^2}{\rho(\rvec)}\right] \mathrm{d}\rvec  =  
4\langle p^2\rangle - 2(2l+1)|m|\langle r^{-2}\rangle \\ 
I_{\pvec} =  \int_{{\mathcal{R}}^3} \left[\frac{|\nabla\Pi(\pvec)|^2}{\Pi(\pvec)}\right] \mathrm{d} \pvec  = 
4\langle r^2\rangle - 2(2l+1)|m|\langle p^{-2}\rangle.
\end{eqnarray}
The above equations can be further recast in following forms,
\begin{eqnarray}
I_{\rvec} = 8\mathcal{E}_{n_r,l}-8\langle v(r)\rangle-2(2l+1)|m|\langle r^{-2}\rangle \\
I_{\pvec} = 8\mathcal{E}_{n_r,l}-8\langle v(p)\rangle-2(2l+1)|m|\langle p^{-2}\rangle .
\end{eqnarray}
where $v(p)$ is the $p$-space counterpart of $v(r)$. 

In case of a CHO, $I$'s in $r$ and $p$ space are expressed analytically as \cite{patil07}, 
\begin{equation}
\begin{aligned} 
I_{\rvec}(\omega)  =\frac{\omega}{\sqrt{2}} I_{\rvec}(\omega=1), \hspace{3mm} \ \ \ \  I_{\pvec}(\omega) = 
\frac{\sqrt{2}}{\omega}I_{\pvec}(\omega=1).
\end{aligned}
\end{equation}
Thus, an increase in $\omega$ leads to rise in $I_{\rvec}(\omega)$ and fall in $I_{\pvec}(\omega)$. However, it is obvious that 
$I_{t} \ (=I_{\rvec} I_{\pvec})$ remains invariant with $\omega$. Throughout the article, for brevity, $I_{\rvec}(\omega=1)$ and 
$I_{\pvec}(\omega=1)$ will be symbolized as $I_{\rvec}$, $I_{\pvec}$ respectively.

When $m=0$, $I_{\rvec}$ and $I_{\pvec}$ in Eqs.~(4), (5) reduce to further simplified forms as below,  
\begin{equation}
\begin{aligned} 
I_{\rvec}  =  4\langle p^2\rangle, \hspace{3mm} \ \ \ \  I_{\pvec} =4\langle r^2\rangle.
\end{aligned}
\end{equation}
It is seen that, at a fixed $n_{r}, l$, both $I_{\rvec}$ and $I_{\pvec}$ provide maximum values when $m=0$, and both of them 
decrease with rise in $m$.  Hence one obtains the following upper bound for $I_t$, 
\begin{equation}
 I_{\rvec} I_{\pvec} \ (=I_t) \leq 16 \langle r^2\rangle \langle p^2\rangle .
\end{equation}
Further adjustment using Eqs.~(6) and (7) leads to following uncertainty relations \cite{romera05}, 
\begin{equation}
\frac{81}{\langle r^2\rangle \langle p^2\rangle} \leq I_{\rvec} I_{\pvec} \leq 16 \langle r^2\rangle \langle p^2\rangle.  
\end{equation}
Therefore, in a central potential, $I$-based uncertainty product is bounded by both upper as well as lower limits. They
are state dependent, varying with $n_{r}, l, m$ quantum numbers.

\section{Result and Discussion}
At the outset, it is instructive to note a few things for ease of discussion. Impact of quantum number $m$ on $I_{\rvec}$ 
and $I_{\pvec}$ of CHO is one of the main objectives of our work; which is performed here for first time. Therefore, to 
ensure a desired accuracy of calculated quantities, a series of background tests were executed. The results in these tables 
are presented up to those points which sustained convergence. $I_{\rvec}$ values are obtained from Eqs.~(4) and (6), whereas 
$I_{\pvec}$ from Eq.~(5). In all occasions, it has been verified that, in both spaces, as $r_c \rightarrow \infty$, 
$I_{\rvec}$ and $I_{\pvec}$ converge to respective FHO limit. The \emph{net} $I$ in $r$ and $p$ spaces are divided
into radial and angular part. But in both $I_{\rvec}$ and $I_{\pvec}$ expressions, angular contribution is normalized to unity. 
Hence, evaluation of all these targeted quantities using only radial part suffices our purpose. The radial wave function 
in $r$ and $p$ spaces depend only on $n_{r}, l$ quantum numbers. Hence, in both space, radial wave function can be obtained
by putting $m=0$ in Eq.~(3). Further, a change in $m$ from \emph{zero} to \emph{non-zero} value will not influence the 
expression of radial wave function in $p$ space. Confinement in isotropic harmonic oscillator is attained by squeezing the 
radial boundary from infinity to a finite region. To realize these, pilot calculations are done for $1s$-$1g$ and $2s$-$2m$ states, 
with $r_c$ varying from 0.1 to 7 a.u. The former set is chosen as they represent nodeless ground states corresponding to various
$l$, whereas, $2s$-$2m$ states are considered to perceive the effect of nodes on $I$ at \emph{non-zero} $m$.
Also note that, we have followed the spectroscopic notation, i.e., the levels are denoted by $n_r+1$ and $l$ values (see, e.g., 
\cite{roy08}). Therefore, $n_r=0$ and $l=4$ signifies $1g$ state. The radial quantum number $n_r$ relates to 
$n$ as $n=2 n_{r}+  l$.

\begingroup            %Table~I  (0s-0g-state)
\squeezetable
\begin{table}
\caption{$\mathcal{E}_{n_r,l} (n_r =0)$ of lowest five circular states of CHO, PISB at five $r_c$. See text for details.}
\centering
\begin{ruledtabular}
\begin{tabular}{l|lllll}
   $l$ & $r_c=0.01$  &  $r_c=0.05$   &  $r_c=0.1^{\dag}$ &   $r_c=0.2$   &   $r_c=0.5^{\dag}$    \\
\hline
\multicolumn{6}{c}{CHO}  \\
\hline
 0   &   49348.02202373  &  1973.92123372 &    493.48163345 &   123.37570844  &   19.77453418  \\
 1   &  100953.64280465  &  4038.14617967 &   1009.53830088 &   252.39159906  &   40.42827649  \\
 2   &  166087.30959293  &  6643.49293120 &   1660.87528919 &   415.22704789  &   66.48975653  \\
 3   &  244155.96823805  &  9753.60193720 &   2441.56211674 &   610.39965899  &   97.72324914  \\
 4   &  334771.55964446  & 13390.86304096 &   3347.71822121 &   836.93939916  &  133.97424683 \\
\hline
\multicolumn{6}{c}{PISB}  \\
\hline
 0   &   49348.022005446   &  1973.92088021   & 493.48022005    &   123.37005501    &     19.73920880  \\
 1   &   100953.64278213   &  4038.14571128   & 1009.53642782   &   252.38410695    &     40.38145711  \\
 2   &   166087.30957134   &  6643.49238285   & 1660.87309571   &   415.21827392    &     66.43492382 \\
 3   &   244155.96821809   &  9753.60153136   & 2441.55968218   &   610.38992054    &     97.66238728  \\
 4   &   334771.55962552   &  13390.86238502  & 3347.71559625   &   836.92889906    &    133.90862385 \\
\end{tabular}
\end{ruledtabular}
\begin{tabbing}
$^\dag${1st three $l$-energies are taken from \cite{roy15}; the remaining ones are freshly generated in this work.}
\end{tabbing}
\end{table}
\endgroup

Exact analytical form of $I_{\rvec}$ and $I_{\pvec}$ in isotropic FHO was given in \cite{romera05},
\begin{equation}
I_{\rvec}(\omega)= 4\omega \left(2n_{r}+l-|m|+\frac{3}{2}\right), \ \ \ 
I_{\pvec}(\omega)=\frac{4}{\omega}\left(2n_{r}+l-|m|+\frac{3}{2} \right). 
\end{equation}
It suggests that, at a fixed $m$, both $I_{\rvec}(\omega), I_{\pvec}(\omega)$ increase as $n_{r}$ and $l$ approach higher values. 
Similarly, for a specific $n_{r}$ and $l$, both $I_{\rvec}(\omega), I_{\pvec}(\omega)$ abate with growth in $|m|$. Influence of 
$\omega$ on 
$I_{\rvec}(\omega)$ and $I_{\pvec}(\omega)$ is quite straightforward. $I_{\rvec}(\omega)$ progresses and $I_{\pvec}(\omega)$ 
reduces with rise of $\omega$. By putting $\omega=1$ in Eq.~(12) one easily gets the expressions for 
$I_{\rvec}$ and $I_{\pvec}$ in a FHO. 
            
\begingroup            %Table~II  (1s-1g-state)
\squeezetable
\begin{table}
\caption{$I_{\rvec}, I_{\pvec}$ of lowest five circular states of CHO and PISB at five $r_c$ at fixed $m$ (0).}
\centering
\begin{ruledtabular}
\begin{tabular}{l|l|lllll}
 System &  $l$ & $r_c=0.01$  &  $r_c=0.05$   &  $r_c=0.1$ &   $r_c=0.2$   &   $r_c=0.5$    \\
\hline
\multicolumn{7}{c}{$I_{\rvec}$}  \\
\hline
     & 0   &   394784.17618984  & 15791.36986976  &   3947.84176    &    986.960440    &   157.913740    \\
     & 1   &   807629.1424372   & 32305.16943736  &   8076.29142    &   2019.072855   &    323.051708232  \\
 CHO & 2   &  1328698.47674344  & 53147.9434496   &   13286.984765  &   3321.7461915  &    531.4794285   \\
     & 3   &  1953247.7459044   & 78028.8154976   &   19532.47745   &   4883.119364   &    781.2991270   \\
     & 4   &  2678172.47715568  & 107126.90432768 &   26781.72476   &   6695.431192   &   1071.269013    \\
\hline
     & 0   &    394784.17604357  &  15791.36704174  &  3947.84176043  &  986.96044010  & 157.91367041   \\
     & 1   &    807629.14225706  &  32305.16569028  &  8076.29142257  & 2019.07285564 &  323.05165690   \\
PISB & 2   &   1328698.47657073  &  53147.93906282  & 13286.98476570  & 3321.74619142 &  531.47939062  \\
     & 3   &   1953247.74574476  &  78028.81225090  & 19532.47745744  & 4883.11936436 &  781.29909829   \\
     & 4   &   2678172.47700420  & 107126.89908016  & 26781.72477004  & 6695.43119251 & 1071.26899080 \\
\hline
\multicolumn{7}{c}{$I_{\pvec}$}  \\
\hline
     & 0   &   0.000113069096  &  0.00282672727 &   0.01130690073 &   0.045227067368  &  0.28253330127  \\
     & 1   &   0.00014984256   &  0.00374606392 &   0.01498424952 &   0.05993660387   &  0.3745037429  \\
CHO & 2   &   0.00017547984   & 0.004386995939 &   0.0175479792  &   0.0701916254    &  0.4386237207  \\
     & 3   &   0.000194769472  & 0.004875545605 &   0.0194769435 &    0.0779075530    &  0.4868660984  \\
     & 4   &   0.000210002530  & 0.005250063214 &   0.0210002501 &    0.0840008284    &  0.5249614675 \\
\hline
     & 0   &   0.000113069096      &  0.00282672741 & 0.01130690966 &  0.04522763864  & 0.28267274151 \\
     & 1   &   0.0001498425609     &  0.00374606402 & 0.01498425609 &  0.05993702437  & 0.374$n$60640235 \\
PISB & 2   &   0.0001754798406     &  0.00438699601 & 0.01754798406 &  0.07019193624  & 0.43869960150 \\
     & 3   &   0.0001947694723     &  0.00486923680 & 0.01947694723 &  0.07790778893  & 0.48692368082 \\
     & 4   &   0.0002100025303     &  0.00525006325 & 0.02100025303 &  0.08400101214  & 0.52500632592 \\
\end{tabular}
\end{ruledtabular}
\end{table}
\endgroup

At first we illustrate the behavior of CHO at $r_c \rightarrow 0$. A careful study exposes that, at small $r_c$, CHO has an 
energy spectrum comparable to that of a particle in a spherical box (PISB). This trend generally holds good for all 
other states as well. A cross-section of eigenvalues ($1s$-$1g$) of lowest five circular states corresponding to $l=0$-4, of 
CHO and PISB, presented in Table~I, at five selected $r_c$, \emph{viz.}, $0.01, 0.05, 0.1, 0.2, 0.5$, supports this fact. 
However this observation should not be misconstrued to conclude that at $r_c \rightarrow 0$, CHO leads to PISB. Because that 
can happen only when both systems have nearly equal kinetic energy as well as potential energy components. This is not 
directly discernible from this table. At this point, it is worthwhile mentioning that, like CHO, PISB is also exactly solvable; 
eigenfunctions are expressible directly in terms of first-order Bessel function, and given as, 
\begin{equation}
J_{l}(Z)=(-1)^{l}Z^{l}\left(\frac{1}{Z}\frac{d}{dZ}\right)^{l}\left(\frac{\sin{Z}}{Z}\right), \\
\end{equation} 
where $Z=\sqrt{\mathcal{E}_{n_{r},l}}\ r$. At the boundary when $r = r_c$, $Z=Z_{n_r,l}$ and $J_{l}(Z_{n_r,l})=0$. Moreover, 
at $r = r_c$, the energy of a $(n_r,l)$ state is expressed as $\mathcal{E}_{n_{r},l}=\frac{Z_{n_r,l}^{2}}{r_{c}^{2}}$ \cite{cohen78}. 
This $J_{l}(Z_{n_r,l})=0$ is a transcendental equation and at a fixed $n_{r},l$, this $Z_{n_{r},l}$ is evaluated by the help of
MATHEMATICA program package. Thus all the PISB energies in lower segment of this table have been obtained following the above 
procedure.  

\begingroup            %Table~III  (1p,1d,1f-state)
\squeezetable
\begin{table}
\caption{$I_{\rvec}, I_{\pvec}$ of $1p, 1d,1f$ states of CHO at six $r_c$, with varying $m$. Results in last column correspond 
to respective FHO values, computed from Eq.~(12). See text for details.}
\centering
\begin{ruledtabular}
\begin{tabular}{l|llllll}
 $|m|$ &  $r_c=0.1$ & $r_c=0.5$  & $r_c=1$   &   $r_c=2$ &   $r_c=7$  &   $r_c=\infty$    \\
\hline
\multicolumn{6}{c}{$I_{\rvec}(1p)$}  \\
%%\hline
 0   &   8076.29142  & 323.051708232 &  80.76619765  & 20.39764116 & 10.0000000000  &  10 \\
 1   &   5736.542528 & 229.424233922 &  57.21792995  & 13.916600785 & 6.0000000000  &   6 \\
\multicolumn{6}{c}{$I_{\pvec}(1p)$}  \\
 0   &   0.014984249523  & 0.374503742927 & 1.491857857098  & 5.577935621669 & 10.00000000  & 10 \\
 1   &   0.01003         & 0.2507         & 0.9980          & 3.975          & 6.0000      &  6 \\
\hline
\multicolumn{6}{c}{$I_{\rvec}(1d)$}  \\
%%\hline
 0   &   13286.984765  &   531.4794285  & 132.8722779   & 33.374501702  & 14.0000000000  & 14  \\
 1   &   10419.849672  &   416.7666252  & 104.09083474  & 25.745587549  & 10.0000000000  & 10  \\
 2   &   7552.714578   &   302.0538220  & 75.3093915    & 18.11667339   &  6.0000000000   &  6  \\
\multicolumn{6}{c}{$I_{\pvec}(1d)$}  \\
 0   &   0.017547979204  &  0.438623720749 & 1.749938940164  & 6.705825222118 & 14.000000000  &  14 \\
 1   &   0.0133333       &  0.33326        & 1.32905         & 5.0595         & 10.00000      &  10 \\
 2   &   0.0091187       &  0.2279         & 0.9081          & 3.4132         &  6.00000      &   6 \\
\hline
\multicolumn{6}{c}{$I_{\rvec}(1f)$}  \\
%%\hline
 0   &   19532.47745   &    781.2991270 &  195.3266196   & 48.95155358 &  18.000000000 &  18 \\
 1   &   16156.649498  &    646.2448005  & 161.48315319  & 40.15669405  & 14.00000000 &  14 \\
 2   &   12780.8215    &    511.1904739 &  127.63968669  & 31.36183453  & 10.00000000 &  10 \\
 3   &   9404.993581   &    376.1361474 &  93.79622020   & 22.566975007 & 6.000000000  &   6 \\
\multicolumn{6}{c}{$I_{\pvec}(1f)$}  \\
 0   &   0.019476943548  & 0.486866098432 & 1.944005783407  & 7.551236509511 & 18.0000000  &  18 \\
 1   &   0.0157907758    & 0.39471708     & 1.57572258      & 6.0995039      & 14.00000   &  14 \\
 2   &   0.012104608     & 0.30256807     & 1.20743938      & 4.6477714      & 10.00000   &   10 \\
 3   &   0.008418440     & 0.2104190      & 0.8391561       & 3.196038       &  6.00000   &   6 \\
\end{tabular}
\end{ruledtabular}
\end{table}
\endgroup

Now the upper portion of Table~II portrays $I_{\rvec}$ of CHO and PISB at five $r_c$ values introduced before; the respective 
$I_{\pvec}$ of two systems are produced in lower segment. There is no literature result available for any of these quantities for 
comparison. It is interesting to note that, for FHO and CHO in a state having $m=0$, $I_{\rvec}$ and $I_{\pvec}$ are directly 
related to expectation values of kinetic and potential energy as;
\begin{equation}
I_{\rvec}(\omega)= 8\langle T \rangle, \ \ \ I_{\pvec}(\omega)=\frac{64}{\omega^{2}}\langle v(r)\rangle. 
\end{equation}
But for a PISB, total energy is solely kinetic energy as the potential energy is zero. However, one can evaluate $I_{\rvec}, 
I_{\pvec}$ for PISB and compare the results with CHO. Because, a pair of systems having same $I_{\rvec}, I_{\pvec}$ for all 
states indicate identical physical and chemical environment. Hence, here we have used $I$ to investigate the characteristics of 
PISB and CHO at $r_c \rightarrow 0$. The table clearly demonstrates that at $r_c \rightarrow 0$, a CHO has comparable $I_{\rvec}, 
I_{\pvec}$ values with that of PISB, thus confirming our presumption that, at $r_c \rightarrow 0$, CHO behaves like a PISB. 
One also notices that, with reduction in $r_{c}$, $I_{\pvec}$ (and also potential energy) approaches \emph{zero}. On the other 
hand, as $r_c$ grows, the separation between $I_{\rvec}$ (also $I_{\pvec}$) values of CHO and PISB tends to extend significantly. 
Moreover, as expected, at $r_c \rightarrow \infty$, CHO reduces to FHO. Elsewhere \cite{Gueorguiev06,laguna14} it has been 
reported that, a 1D CHO may be treated as a two-mode system; at smaller (approaching zero) and larger (tending infinity) confinement 
lengths it behaving like a particle in a box and an FHO respectively. Here, also we observe analogous behavior. At 
$r_c \rightarrow 0$ and $\infty$, CHO leads to PISB and a 3D isotropic FHO respectively. We also note that, larger the value of 
$\omega$ 
higher will be $\langle v(r)\rangle$; as a consequence CHO is more prone to FHO in such a case. On the contrary, lesser the 
$\omega$, $\langle v(r)\rangle$ is smaller and CHO, in that occasion, resembles more to a PISB. Therefore at a fixed $r_c$, by 
modulating $\omega$ values one can investigate properties of all three systems starting from PISB to FHO through CHO.  

So far, we have been discussing about the limiting trend of CHO. Now we turn focus on to its behavior at intermediate
$r_c$ region. For that, at first, the dependence of $I_{\rvec}, I_{\pvec}$ on quantum number $n_r$ is recorded in Table~S1 of 
Supplementary Material (SM). It tabulates these quantities for lowest five $n_r$ (0-4) at six representative $r_c$. This plainly
implies that, at fixed $m,l$ and $r_c$, both $I_{\rvec}, I_{\pvec}$ in CHO get incremented as $n_{r}$ assumes higher values. 
Henceforth, the role of $n_{r}$ on these measures is not discussed any further.   

\begin{figure}                         %%%Fig. 1, CHO
\begin{minipage}[c]{0.35\textwidth}\centering
\includegraphics[scale=0.50]{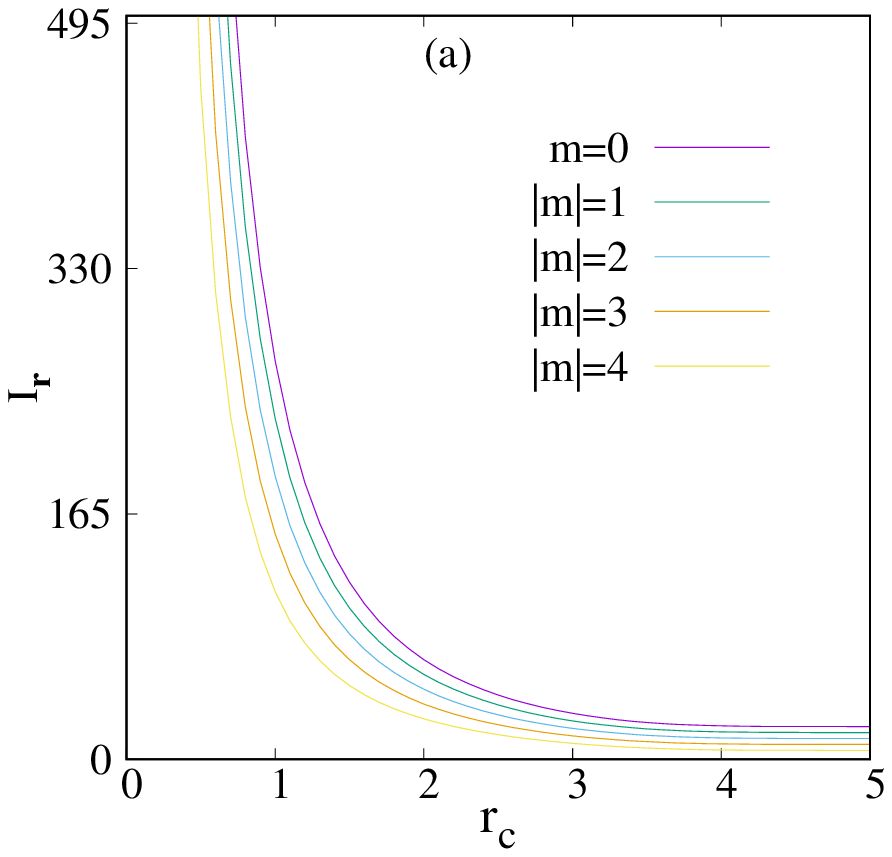}
\end{minipage}%
\vspace{1mm}
\begin{minipage}[c]{0.35\textwidth}\centering
\includegraphics[scale=0.50]{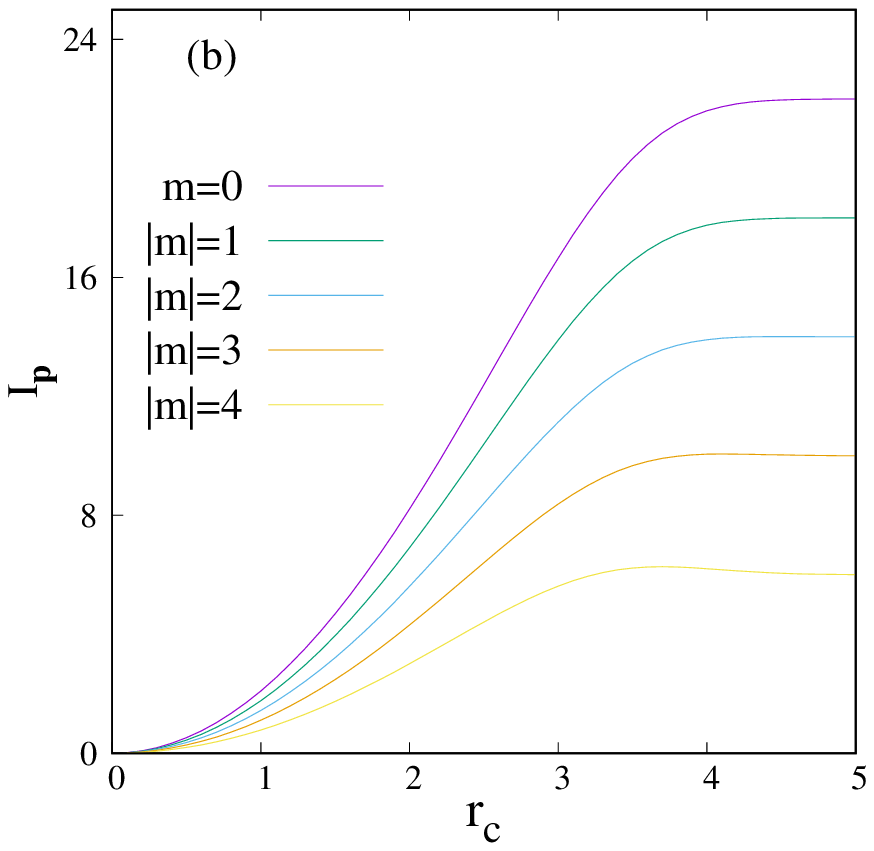}
\end{minipage}%
\caption{Variation of $I_{\rvec}$ and $I_{\pvec}$ in CHO, with $r_c$, for all allowed $|m|$ values of $1g$ state, in panels 
(a) and (b) respectively. See text for details.}
\end{figure}

Now, in order to get a better picture of the effect of magnetic quantum number, Table~III gives $I_{\rvec}, I_{\pvec}$ of CHO 
for lowest three circular states having $l \neq 0$, i.e., $1p$-$1f$ respectively, for all possible $m$. Likewise, Fig.~1 displays 
these for $1g$ state in panels (a) and (b), for all permitted $|m|$ at 6 carefully selected $r_c$ ($0.1, 0.5, 1, 2, 7, \infty$). 
Once again, no literature works exist along these lines; hopefully they would be useful for future referencing. 
It is to be noted that, the quantities in last column at $r_c = \infty$ are given from Eq.~(12) considering $\omega=1$. 
For this special $\omega$, Eq.~(12) dictates that, $I_{\rvec}=I_{\pvec}$ for FHO. One notices that, behavior of $I_{\rvec}$ and 
$I_{\pvec}$ in CHO is always in consonance with FHO; a thorough analysis suggests identical patterns. In general, $I_{\rvec}$ 
decays monotonically while $I_{\pvec}$ progresses with advancement of $r_c$; this is found to hold good for all $|m|$. This is 
to be expected, as a growth in $r_c$ facilitates delocalization in $r$ space and localization in $p$ space. At a certain $r_c$ and 
fixed quantum numbers $n_{r}$ and $m$, $I_{\rvec}$ tends to grow with $l$; this is consistent to what is observed from Eq.~(12) in FHO. 
Because, in both CHO and FHO, kinetic energy gains with $n_{r},l$. Similarly, at a fixed $n_{r}, l$, in both CHO and FHO, 
$I_{\rvec}$ lessens as one descends down the table (increment in $|m|$). One striking fact is that, for all these three states 
considered, both CHO and FHO portray similar pattern with respect to $m$. Next, $I_{t}$ for $1s$-$1f$ states of CHO are offered
in Table~S2 of SM. It seems to show a propensity towards $I_{\rvec}$ in the behavioral pattern. In all instances, the lower and
upper bounds governed by Eq.~(11) is satisfied. For an arbitrary state distinguished by quantum numbers $n_r,l$, variation in 
$I_{\rvec}, I_{\pvec}$ with $r_c$ in a CHO preserves same qualitative orderings for various $m$ (general nature of the plots
remain unchanged) as depicted in these figures. This has been verified in multiple occasions, which are not reported here to save 
space. As usual, in all cases, they all eventually approach their respective CHO-limit at some sufficiently large $r_c$, which 
varies from state to state. 

\begingroup            %Table~IV
\squeezetable
\begin{table}[tbp]
\caption{$I_{\rvec}, I_{\pvec}$ for $2p$-$2m$ ($|m|=1$) states in CHO, at six selected $r_c$. See text for details.}
\centering
\begin{ruledtabular}
\begin{tabular}{l|llllll}
 $l$ &  $r_c=0.1$  & $r_c=0.5$  &   $r_c=1$ &   $r_c=2$   &   $r_c=7$ &   $r_c=\infty$   \\
\hline
\multicolumn{7}{c}{$I_{\rvec}$}  \\
\hline
 1   & 19544.75146    & 781.7758499    & 195.3902144   & 48.61008407   & 14.000000000    & 14  \\
 2   & 28201.56829    & 1128.0488516   & 281.9599554   & 70.272141773  & 18.000000000   & 18   \\
 3   & 37976.00550    & 1519.027385    & 379.70865634  & 94.73419843   & 22.0000000000    & 22  \\
 4   & 48838.86407    & 1953.5428786   & 488.3419262   & 121.91517166  & 26.0000000000    & 26  \\
 5   & 60766.49285    & 2430.649110    & 607.62258270  & 151.75456390  & 30.0000000001   & 30  \\ 
 6   & 73739.4732     & 2949.5692932   & 737.3562778   & 184.204055587 & 34.0000000743   & 34  \\
 7   & 87741.58879    & 3509.6547640   & 877.38084537  & 219.22364350  & 38.00000037     & 38  \\
 8   & 102759.07974   & 4110.3551463   & 1027.5587357  & 256.77944910  & 42.0000017797    & 42  \\
 9   & 118780.106607  & 4751.196872    & 1187.77161130 & 296.84230573  & 46.0000074341   & 46  \\
\hline
\multicolumn{7}{c}{$I_{\pvec}$}  \\
\hline
 1   & 0.01221      & 0.3054        & 1.222        & 4.917       & 14.000      & 14  \\
 2   & 0.0133333    & 0.333337      & 1.33356      & 5.34750     & 18.00000    & 18  \\
 3   & 0.014439157  & 0.36097833    & 1.4438781    & 5.7743588   & 21.999997   & 22 \\
 4   & 0.0154743767 & 0.38685628    & 1.54723822   & 6.1782594   & 26.0000000  & 26  \\
 5   & 0.016426161  & 0.4106495     & 1.6423270    & 6.55303527  & 30.000000   & 30  \\ 
 6   & 0.017296672  & 0.4324115849  & 1.7293332625 & 6.898114452 & 34.00000    & 34  \\
 7   & 0.0180927445 & 0.4523131126  & 1.808922468  & 7.21517631  & 38.00000    & 38  \\
 8   & 0.0188222240 & 0.470550071   & 1.88186836   & 7.50667165  & 41.9999991  & 42  \\
 9   & 0.019492648  & 0.487310795   & 1.94891804   & 7.775182    & 45.99998    & 46  \\
\end{tabular}
\end{ruledtabular}
\end{table}
\endgroup

Now to understand the effect of $l$ on $I_{\rvec}, I_{\pvec}$, we offer Table~IV, where these are listed for $l=1-9$ states 
having $|m|=1$ and $n_{r}$ corresponding to 1, at same six chosen $r_c$'s of previous table. As earlier, here too, no reference 
work is known to us. The last column again has same 
significance as Table~III. In accordance with Eq.~(12), here also for $2l$ states, $I_{\rvec}=I_{\pvec}$ in FHO. Dependence of 
$I_{\rvec}, I_{\pvec}$ of CHO on $l$ parallels our observation in FHO. At a given $n_{r}, m$ and $r_c$, they both get incremented 
with rise of $l$ in CHO and FHO. This may occur presumably because that, as $l$ progresses, the density gets increasingly concentrated. 
Therefore, at fixed $n_{r}, m$, a state with higher $l$ experiences greater oscillation. Thus all our foregoing discussion leads 
to a general fact that, the qualitative variations of $I_{\rvec}$ with all three quantum numbers remain quite similar to that of 
$I_{\pvec}$ in a CHO; also the trends in CHO and FHO are analogous. It may be appropriate to mention a parallel work 
\cite{mukherjee17} along this line for free and confined H atom inside an impenetrable spherical enclosure. One finds several 
significant deviations in the variation pattern between two systems there. Here we mention two of the most interesting facts, which 
are in direct contrast with a CHO, e.g., $I_{\rvec}$ remains invariant with respect to changes in $l$, whereas under confinement, it 
reduces with $l$ (at fixed $n,m$). Besides, for a given state having fixed $n,m$, $I_{\pvec}$ progresses when the atom is 
compressed, whereas declines in a free H atom. 

\begin{figure}                         %%%Fig. 2, CHO
\begin{minipage}[c]{0.30\textwidth}\centering
\includegraphics[scale=0.45]{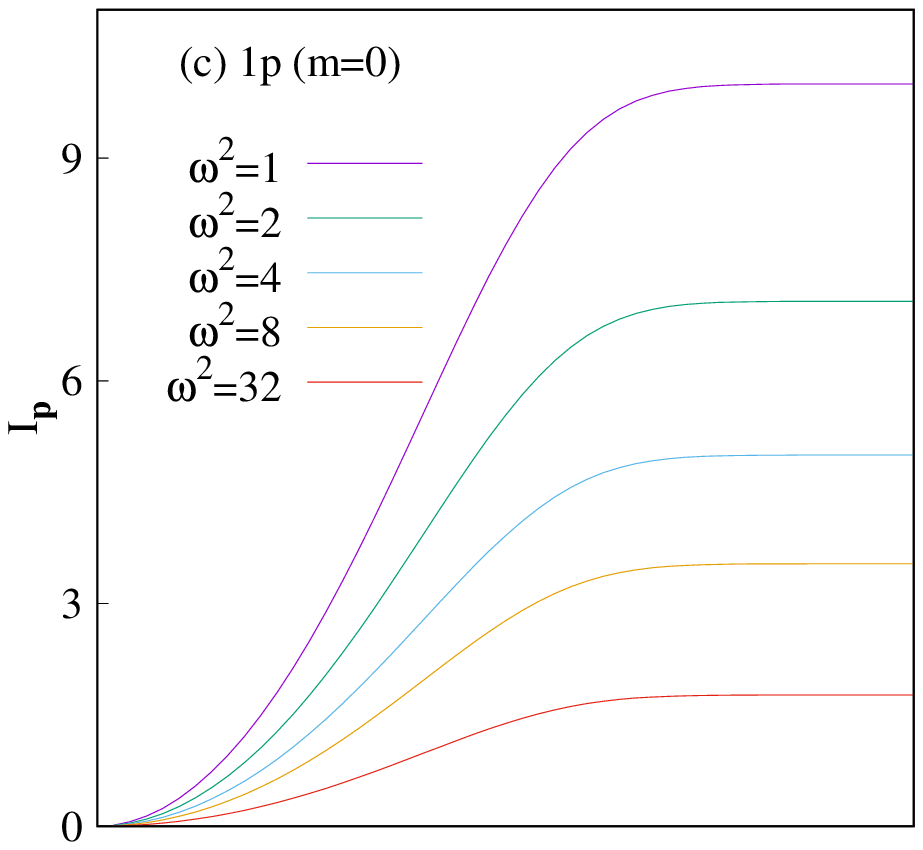}
\end{minipage}%
\vspace{1mm}
\begin{minipage}[c]{0.30\textwidth}\centering
\includegraphics[scale=0.45]{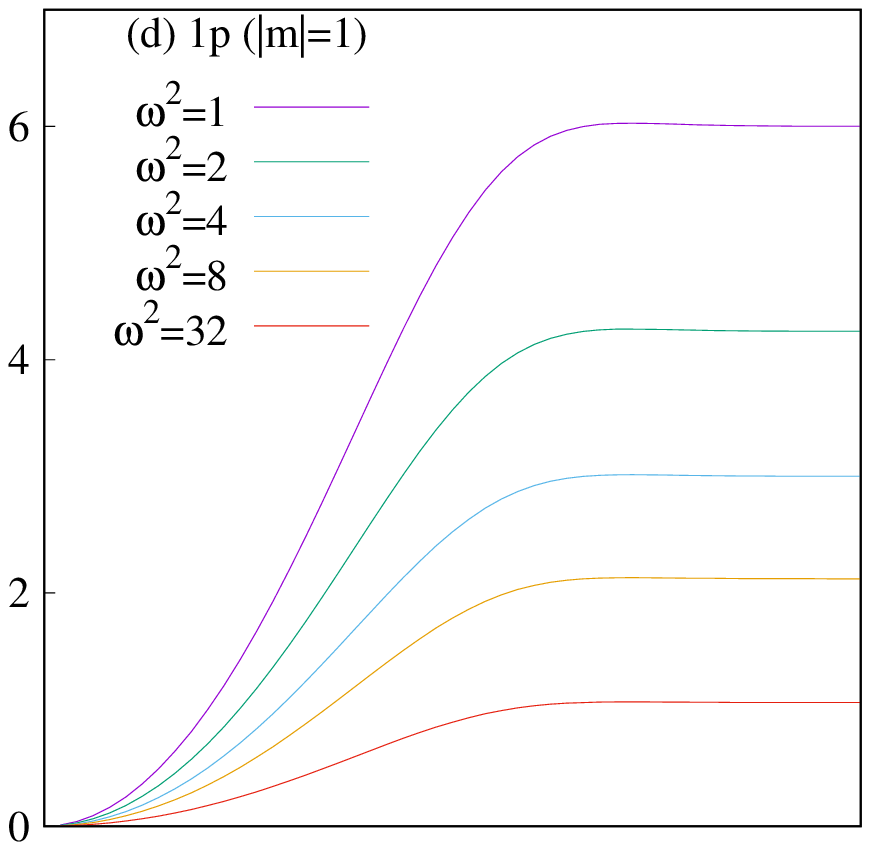}
\end{minipage}%
\vspace{1mm}
\hspace{0.2in}
\begin{minipage}[c]{0.32\textwidth}\centering
\includegraphics[scale=0.48]{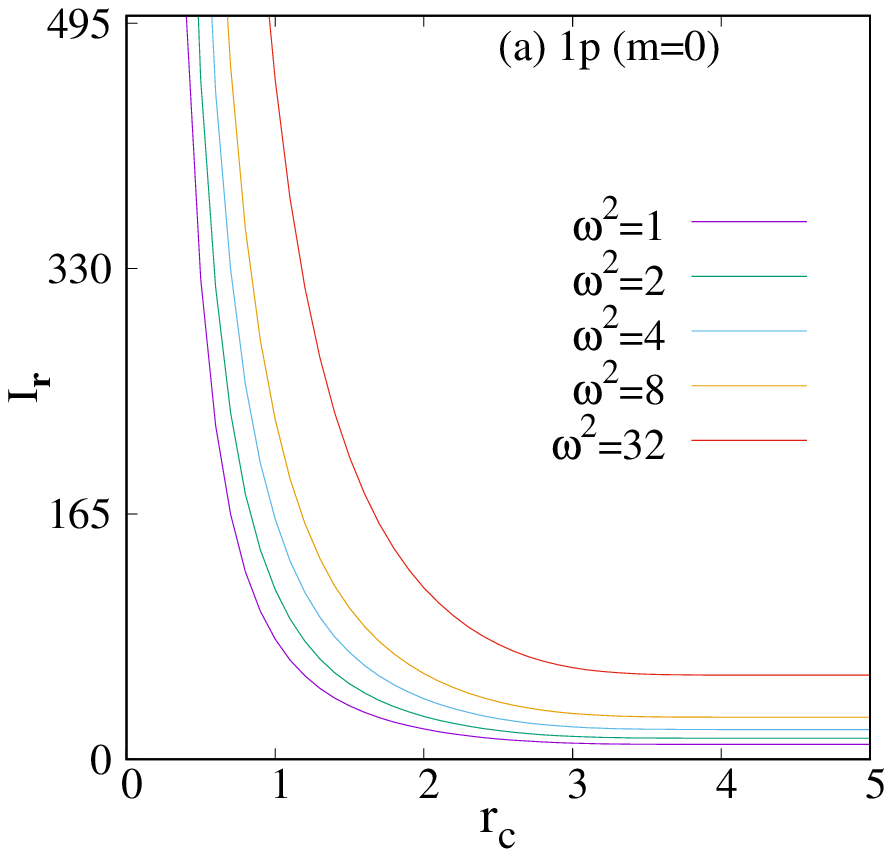}
\end{minipage}%
\begin{minipage}[c]{0.32\textwidth}\centering
\includegraphics[scale=0.48]{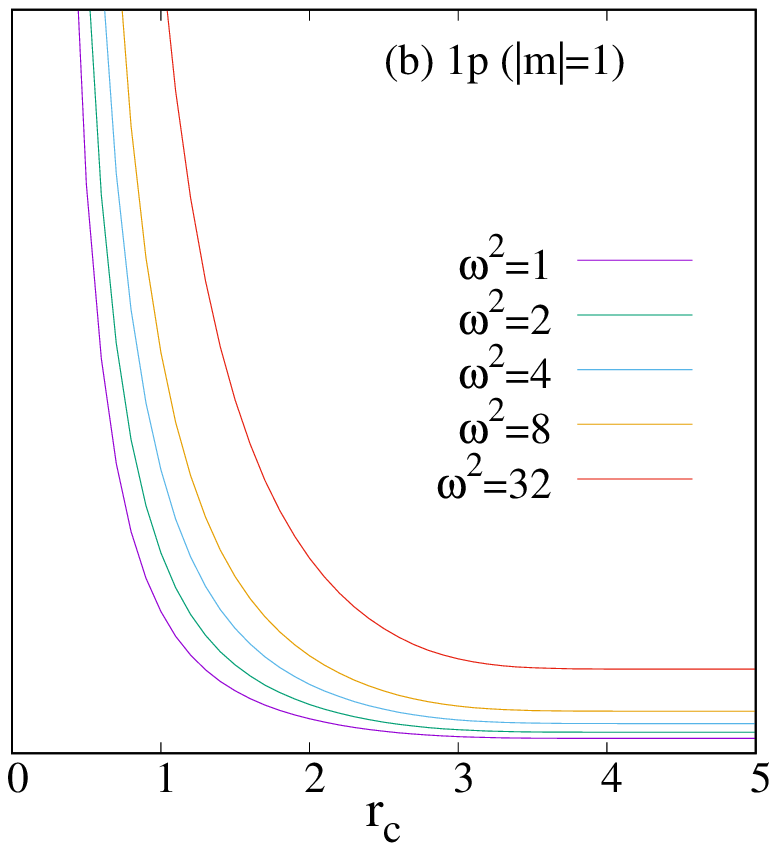}
\end{minipage}%
\caption{Plots of $I_{\rvec}$ and $I_{\pvec}$ in bottom ((a), (b)) and top ((c),(d)) panels in CHO with $r_c$, of 
$1p$ state at five selected $\omega^{2} (1, 2, 4, 8, 32)$. Left, right columns correspond to $|m|=0$, 1 respectively.}
\end{figure}

Before concluding, a few words may be devoted to the influence of $\omega$ on $I_{\rvec}, I_{\pvec}$. In order to probe this, 
Fig.~2 depicts 
plots of $I_{\rvec}$ and $I_{\pvec}$ against $r_c$, for 5 selected $\omega^{2}$ ($1, 2, 4, 8, 32$), in bottom ((a), (b)) and 
top ((c), (d)) panels. These are given for $1p$ state; left and right panels characterize $|m|=0$ and 1 respectively. Evidently 
at a given $r_c$, $I_{\rvec}$ grows and $I_{\pvec}$ decays with increment in $\omega$. At a fixed $\omega$, dependence of these 
measures on $n_{r},l,m$ is similar to that in FHO. As $\omega$ goes up, there is more localization, hence compactness in 
single-particle density enhances with oscillation strength.

\section{Future and Outlook}
Benchmark values of $I_{\rvec}, I_{\pvec}, I_t$ are offered in a CHO, for both $l=0$ and $l \neq 0$ states with special 
emphasis on the latter. Representative results are provided for $1s$-$1g$ as well as $2s$-$2m$ states. A CHO may be considered as 
a two-mode system; at $r_c \rightarrow 0$ it behaves as a PISB and at $r_c \rightarrow \infty$ it reduces to an FHO. Effect of $m$ 
on these measures has been analyzed in detail. To the best of our knowledge, such an examination in CHO is pursued here as a 
first case. With progress in $r_c$, $I_{\rvec}$ falls and $I_{\pvec}$ rises. These are compared and contrasted with FHO results; 
in all aspects they show analogous trends. Further, their changes with respect to $\omega$ is also registered. This work suggests 
that, $I$ may be employed to predict various properties of physically important systems. Therefore, further inspection of $I$ 
in other free and confinement situations may be worthwhile and could be taken up in future. 

\section{Acknowledgement}
Financial support from DST SERB, New Delhi, India (sanction order: EMR/2014/000838) is gratefully acknowledged. NM thanks DST 
SERB, New Delhi, India, for a National-post-doctoral fellowship (sanction order: PDF/2016/000014/CS).

\end{document}

%%%%%%%%%%%%%%%%%%%%%%%%%%%%%%% SUPPLYMENTARY MATERIAL %%%%%%%%%%%%%%%%%%%%%%%%%%%%%%%%%%%%%%%

\pagebreak
\widetext

\begin{center}
\textbf{\large Supplemental Materials: Fisher information in confined isotropic harmonic oscillator.}
\end{center}

\setcounter{page}{1}
\setcounter{figure}{0}
\setcounter{table}{0}

\makeatletter
\renewcommand{\thefigure}{S\arabic{figure}}
\renewcommand{\thetable}{S\arabic{table}}

\vspace{1mm}

\begingroup            %Table~I  (1s-1g-state)
\squeezetable
\begin{table}[h]
\caption{$I_{\rvec}, I_{\pvec}$ for lowest five $n_r$ ($l=0$) values at six different $r_c$. See text for details.}
\centering
\begin{ruledtabular}
\begin{tabular}{l|llllll}
   $n_{r}$ & $r_c=0.1$  &  $r_c=0.5$   &  $r_c=1$ &   $r_c=2$   &   $r_c=7$  & $r_c=\infty$    \\
\hline
\multicolumn{7}{c}{$I_{\rvec}$}  \\
\hline
 0   &   3947.84176  &  157.9137401 &    39.48285935 &  10.130828577  &  6.00000000  & 6  \\
 1   &  15791.3670  &  631.654662 &   157.91245186   &  39.42241043   & 14.00000000  & 14 \\
 2   &  35530.57584  &  1421.2230213 &   355.3049651 &  88.77709457   & 22.00000000  & 22 \\
 3   &  63165.46816  &  2526.6187189 &   631.6541846 &  157.88183967  & 30.00000000  & 30  \\
 4   &  98696.0440  & 3947.8417551 &   986.960106    &  246.718681361 & 38.00000000  & 38 \\
\hline
\multicolumn{7}{c}{$I_{\pvec}$}  \\
\hline
 0   &  0.0113069007    &  0.2825333012  & 1.1217967676   & 3.9877029335    & 6.0000000      & 6  \\
 1   &  0.01282672989   & 0.32070687721 & 1.28512015257   & 5.25470219890   & 14.0000000     & 14 \\
 2   &  0.0131081767    & 0.3277292039  &  1.3124046596   &  5.34276239849  & 22.00000000    & 22 \\
 3   &  0.0132066828    & 0.3301825760   & 1.3216621568   & 5.3463257844    & 30.00000000    & 30 \\
 4   &  0.0132522770    & 0.33131731951   & 1.3258940221   & 5.3437134181   & 38.00000000    & 38 \\
\end{tabular}
\end{ruledtabular}
\end{table}
\endgroup

\begingroup            %Table~III  (1p,1d,1f-state)
\squeezetable
\begin{table}[h]
\caption{$I_{t}$ for $1p$-$1f$ $(n_{r}=0)$ orbitals at six different $r_c$. See text for details.}
\centering
\begin{ruledtabular}
\begin{tabular}{l|llllll}
 $|m|$ &  $r_c=0.1$ & $r_c=0.5$  & $r_c=1$   &   $r_c=2$ &   $r_c=7$  &   $r_c=\infty$    \\
\hline
\multicolumn{7}{c}{$I_{t}(1p)^{a}$}  \\
\hline
 $0^{\dag}$   &   121.01515 & 120.98286 &  120.491051  & 113.776614 & 100.00000000  &  100 \\
 1            &   57.53752 & 57.5166 &  57.1034     & 55.318     & 36.0000  &   36 \\
\hline
\multicolumn{7}{c}{$I_{t}(1d)^{b}$}  \\
\hline
 $0^{\dag}$  & 233.15973   &  233.117506  & 232.518248 & 223.80357 & 196.000000000  & 196  \\
 1           & 138.93098   &  138.89164  & 138.3419   & 130.2598  & 100.00000    & 100  \\
 2           & 68.870938   &   68.8380  & 68.3884    &  61.8358  & 36.00000    &  36  \\
\hline
\multicolumn{7}{c}{$I_{t}(1f)^{c}$}  \\
\hline
$0^{\dag}$  & 380.432960   &  380.388051   & 379.716077 & 369.644758 & 324.0000000 &  324 \\
 1          & 255.126029   &  255.083860   & 254.452650 & 244.935911 & 196.00000 &  196 \\
 2          & 154.706834   &  154.669915   & 154.117184 & 145.762637 & 100.00000 &  100 \\
 3          & 79.1753741   &  79.1461919   & 78.7096703 & 72.1249096 &  36.00000 &   36 \\
\end{tabular}
\end{ruledtabular}
\begin{tabbing}
{$^\dag$ These also correspond to upper bounds, given in Eq.~(10).} \\
{$^a$ Lower bounds Eq.~(11), for $1p$ at 6 $r_c$ are: 10.70940, 10.71226, 10.75598, 11.39074, 12.96, 12.96.} \\
{$^b$ Lower bounds Eq.~(11), for $1d$ at 6 $r_c$ are: 5.55842, 5.559428, 5.573756, 5.79079, 6.61224, 6.61224.} \\
{$^c$ Lower bounds Eq.~(11), for $1f$ at 6 $r_c$ are: 3.40664, 3.40704, 3.41307, 3.506068, 4, 4.}   
\end{tabbing}
\end{table}
\endgroup